\documentclass[prc,showpacs,showkeys]{revtex4}
\usepackage{epsfig}
\begin{document}
%\begin{frontmatter}
\title{In-medium omega meson mass and quark condensate in a Nambu Jona-Lasinio model constrained by recent experimental data}
\author{R. Huguet, J.C. Caillon and J. Labarsouque}
\address{Universit\'{e} Bordeaux 1 ; CNRS/IN2P3 ; \\Centre d'Etudes Nucl\'{e}aires de Bordeaux-Gradignan, UMR 5797 \\Chemin du Solarium, BP120, 33175 Gradignan, France}
\begin{abstract}
We have determined the relation between the in-medium $\omega $ meson mass and quark condensate in the framework of a Nambu Jona-Lasinio model constrained by some recent experimental data on the meson properties in nuclei. In addition to the usual four-quark
interactions, we have included eight-quark terms in the Lagrangian. The parameters of this model have been determined using the meson properties in the vacuum as well as in the medium. More particularly, we have constrained both the in-medium pion decay constant
to the value measured in experiments on deeply bound pionic atoms and the in-medium $\omega $ meson mass to the experimental value obtained either by the TAPS collaboration
or by the E325 experiment at KEK. Our results are
compared to several scaling laws and in particular to that of Brown
and Rho.\\

\end{abstract}

%\end{keyword}
%\end{frontmatter}
\maketitle
%\date{}

\section{Introduction}

These last years, much attention has been focussed on the modification of
hadron properties in nuclear environment and more particularly in the sector
of the light vector mesons. The hope is that this modification could shed some light on prominent features of QCD at low energy. In particular, the knowledge of the dependence
of the in-medium vector meson mass on the quark condensate is essential to a
better understanding of the role played by the chiral structure of the QCD
vacuum.

Experimentally, an indirect indication of the modification of hadron
properties in the medium has been provided by the dilepton production
measurements in relativistic heavy-ion collisions, like for example,
experiments from CERES\cite{cer} and HELIOS\cite{hel} collaborations. However,
the interpretation in terms of a reduction of the $\rho $ mass is still controversial.
Recently, new experiments using proton-induced nuclear reactions\cite{nar},
or $\gamma -A$ reactions\cite{tap} have provided a more clear experimental signature of the
in-medium modifications of the $\omega $ mesons. In particular, the
modification in nuclei of the $\omega $ meson has been investigated in
photoproduction experiments by the TAPS collaboration\cite{tap} and its mass
was found to be $m_{\omega }^{*}=722_{-4}^{+4}$ (stat)$_{-5}^{+35}$(syst)
MeV at 0.6 times the saturation density of nuclear matter. The same order of
magnitude, a 9\% decrease of the in-medium $\omega $ mass at saturation, has
been observed by Naruki et al.\cite{nar} in 12 GeV proton-nucleus reactions
(E325/KEK).

On the other hand, experimental indications of the in-medium modification of
the quark condensate, $\left\langle \overline{q}q\right\rangle $, can be
obtained, for example, in experiments on deeply bound pionic atoms. Indeed,
by deducing the isovector $\pi N$ interaction parameter in the
pion-nucleus potential from the binding energy and 
width of deeply bound 1s states of $\pi ^{-}$ in
heavy nuclei, the in-medium pion decay constant, $f_{\pi }^{*}$, can be
extracted\cite{gei,suz}. The quark condensate is then connected to $%
f_{\pi }^{*}$ through the Gell-Mann-Oaks-$%
%TCIMACRO{\func{Re}}
%BeginExpansion
\mathop{\rm Re}%
%EndExpansion
$nner relation. The observed enhancement of the isovector $\pi N$ interaction parameter over the free $\pi N$
value indicates a reduction of the pion decay constant in the
medium which was found\cite{suz} to be $f_{\pi }^{*2}/f_{\pi }^{2}=0.64$ at saturation density of nuclear matter.

>From a theoretical point of view, starting from the assumption of Harada and
Yamawaki\cite{h.y} on the ''vector manifestation'' of chiral symmetry in
which a hidden local symmetry theory is matched to QCD, Brown and Rho
proposed\cite{br.} that, up to the saturation density, the vector meson mass
in medium, $m_{V}^{*}$, scales according to the approximative relation : $%
m_{V}^{*}/m_{V}\sim \left[ \left\langle \overline{q}q\right\rangle
/\left\langle \overline{q}q\right\rangle _{0}\right] ^{1/2}$ (where $%
\left\langle \overline{q}q\right\rangle $ and $\left\langle \overline{q}%
q\right\rangle _{0}$ are respectively the in-medium and vacuum quark
condensates). In quite different frameworks, like, for example, in finite
density QCD sum rule calculations\cite{hat,asa,koi,jin,kli} or in the Nambu
Jona-Lasinio model (NJL)\cite{b.m}, the relation between the in-medium
vector meson mass and quark condensate is not so clear
and thus more complicated to handle.

The recent experimental data, like those previously mentioned, should
provide stringent tests for the models and for the relation between the
in-medium $\omega $ meson mass and quark condensate. An indication on the
consequences of these new constraints could be obtained by enforcing them in
quark models incorporating the most prominent features of QCD. In this
context, the NJL model\cite{njl} appears as a good candidate since it allows
a dynamical description of both the breaking of chiral symmetry
and of the modification of the in-medium $\omega $ meson mass.

In this work, we have determined the dependence of the in-medium $\omega $
meson mass on the quark condensate in a NJL model constrained by
in-medium meson properties in accordance with recent
experimental data. In addition to the usual four-quark interactions, we have included eight-quark terms in the NJL Lagrangian. The
parameters of this model have been determined using the meson properties in
the vacuum as well as in the medium through the pion decay constant and $%
\omega $ meson mass. More particularly, the in-medium pion decay constant
has been constrained by the value obtained in an experiment on deeply bound
pionic atoms\cite{suz} and we have fixed the in-medium $\omega $ meson mass to the 
experimental values determined by the TAPS collaboration\cite{tap} or by the 
E325 experiment at KEK\cite{nar}. These in-medium changes of meson properties arise from dynamical chiral symmetry restoration at the quark mean-field-RPA level as well as from more complicated quark-gluon excitations usually parametrized in terms of many-body hadronic interactions. Considering the importance of the role played by the chiral structure of the QCD vacuum, concerning the $\omega $ meson mass and the pion decay constant, we made the assumption that, to leading order in nuclear density, the main contribution comes from dynamical chiral symmetry restoration at the quark mean-field-RPA level. Our results will be compared to several scaling laws and in particular to that of Brown and Rho.

\section{Formalism}

We consider the following chirally invariant two-flavor NJL Lagrangian\cite
{njl} up to eight-quark interaction terms :

\begin{eqnarray}
{\cal \ L} &=&\overline{q}\left[ i\gamma _{\mu }\partial ^{\mu
}-m_{0}\right] q+g_{1}\left[ (\overline{q}q)^{2}+(\overline{q}i\gamma _{5}%
{\bf \tau }q)^{2}\right] -g_{2}(\overline{q}\gamma _{\mu }q)^{2}  \label{njl}
\\
&&+g_{3}\left[ (\overline{q}q)^{2}+(\overline{q}i\gamma _{5}{\bf \tau }%
q)^{2}\right] (\overline{q}\gamma _{\mu }q)^{2}+g_{4}\left[ (\overline{q}%
q)^{2}+(\overline{q}i\gamma _{5}{\bf \tau }q)^{2}\right] ^{2}, \nonumber
\end{eqnarray}

\noindent where $q$ denotes the quark field with two flavor ($N_{f}=2$) and
three color ($N_{c}=3$) degrees of freedom and $m_{0}$ is the diagonal
matrix of the current quark masses (here in the isospin symmetric case). The
second and third terms of Eq.\ref{njl} represent local four-quark
interactions while those proportional to $g_{3}$ and $g_{4}$ are
eight-quark interactions. Let us recall that, in two-flavor models, the t'Hooft six fermion interaction term can be rewrited in terms of the four-quark interactions considered here\cite{kle}. We have not considered the term $(%
\overline{q}\gamma _{\mu }q)^{4}$ since, as in the nucleonic NJL model\cite{mis},
it leads to a violation of the causality at high density.

The Dirac equation for a quark in mean-field approximation is given by :

\begin{equation}
\left[ i\gamma _{\mu }\partial ^{\mu }-m_{0}-2g_{2}\gamma _{0}\left\langle 
\overline{q}\gamma _{0}q\right\rangle +2g_{1}\left\langle \overline{q}%
q\right\rangle +2g_{3}\left\langle \overline{q}q\right\rangle \left\langle 
\overline{q}\gamma _{0}q\right\rangle ^{2}+4g_{4}\left\langle \overline{q}%
q\right\rangle ^{3}\right] q=0,  \label{dir}
\end{equation}

\noindent which defines a dynamical constituent-quark mass :

\begin{equation}
m=m_{0}-2g_{1}\left( 1+\frac{g_{3}N_{f}^{2}N_{c}^{2}\rho _{B}^{2}}{4g_{1}}+%
\frac{2g_{4}}{g_{1}}\left\langle \overline{q}q\right\rangle ^{2}\right)
\left\langle \overline{q}q\right\rangle ,  \label{gap}
\end{equation}

\noindent generated by a strong scalar interaction of the quark with the
Dirac vacuum. In the gap equation (Eq.\ref{gap}), the quark condensate $%
\left\langle \overline{q}q\right\rangle $ can be written as :

\begin{equation}
\left\langle \overline{q}q\right\rangle =-i\int \frac{d^{4}k}{\left( 2\pi
\right) ^{4}}{\tt Tr}S(k),  \label{qqbs}
\end{equation}
where here Tr denotes traces over color, flavor and spin. In Eq.\ref{qqbs}, $%
S(k)$ represents the in-medium quark propagator defined as :

\begin{equation}
S(k)=\frac{1}{\gamma _{\mu }k^{\mu }-m+i\varepsilon }+i\pi \frac{\gamma
_{\mu }k^{\mu }+m}{E_{k}}\delta \left( k_{0}-E_{k}\right) \theta \left(
k_{F}-\left| {\bf k}\right| \right) ,  \label{pro}
\end{equation}

\noindent where $E_{k}=\sqrt{{\bf k}^{2}+m^{2}}$ and $k_{F}$ is the quark
Fermi momentum. The baryonic density is related to the total quark density
by $\rho _{B}=\frac{1}{3}\rho _{q}$. The quark condensate is divergent due
to the loop integrals and requires an appropriate regularization procedure.
As many authors\cite{b.m,bub}, we introduce a three-momentum cutoff $\Lambda 
$ which has the least impact on medium parts of the regularized integrals,
in particular at zero temperature\cite{bub}. Thus, after the regularization
procedure, the quark condensate is given at each density by :

\begin{equation}
\left\langle \overline{q}q\right\rangle =-\frac{N_{f}N_{c}}{\pi ^{2}}%
\int_{k_{F}}^{\Lambda }\frac{mk^{2}dk}{E_{k}}.  \label{qqb}
\end{equation}

As usual, the $\omega$ meson mass is obtained by solving the Bethe-Salpeter
equation in the quark-antiquark channel. First, we define the
quark-antiquark polarization operator in the $\omega $ channel by :

\begin{eqnarray}
\Pi _{\omega }^{\mu \nu }(q^{2}) &=&-i\int \frac{d^{4}p}{\left( 2\pi \right)
^{4}}{\tt Tr}\left[ i\gamma ^{\mu }iS(p+q/2)i\gamma ^{\nu }iS(p-q/2)\right]
\label{pola} \\
&=&\left( -g^{\mu \nu }+\frac{q^{\mu }q^{\upsilon }}{q^{2}}\right) \Pi
_{\omega }(q^{2}),  \nonumber
\end{eqnarray}
where $S(k)$ is given by Eq.\ref{pro}. Using the same regularization
procedure as for the quark condensate, we obtain : 
\begin{equation}
\Pi _{\omega }(q^{2})=\frac{N_{f}N_{c}}{12\pi ^{2}}q^{2}%
\int_{4(p_{F}^{2}+m^{2})}^{4(\Lambda ^{2}+m^{2})}\frac{\sqrt{1-4m^{2}/\kappa
^{2}}}{q^{2}-\kappa ^{2}}\left( 1+\frac{2m^{2}}{\kappa ^{2}}\right) d\kappa
^{2}.  \label{polw}
\end{equation}

\noindent The in-medium $\omega $ meson mass, $m_{\omega }^{*}$, is then
determined by the pole structure of the $T$-matrix, i.e. by the condition :

\begin{equation}
1-2\left( g_{2}-g_{3}\left\langle \overline{q}q\right\rangle ^{2}\right) \Pi
_{\omega }(q^{2}=m_{\omega }^{*2})=0.  \label{mw}
\end{equation}

We also need the pion mass and decay constant to adjust the model
parameters. In the pseudo-scalar channel, the polarization reads : 
\begin{equation}
\Pi _{\pi }(q^{2})=\frac{\left\langle \overline{q}q\right\rangle }{m}%
+N_{c}N_{f}q^{2}I(q^{2}),  \label{ppi}
\end{equation}
with

\begin{equation}
I(q^{2})=\frac{1}{8\pi ^{2}}\int_{4(\Lambda ^{2}+m^{2})}^{4(p_{F}^{2}+m^{2})}%
\frac{1}{q^{2}-\kappa ^{2}}\sqrt{1-\frac{4m^{2}}{\kappa ^{2}}}d\kappa ^{2}.
\label{iq2}
\end{equation}
The in-medium $\pi $ meson mass, $m_{\pi }^{*}$, and decay constant, $f_{\pi
}^{*}$, are then given respectively by :

\begin{equation}
1-2\left( g_{1}+\frac{g_{3}N_{f}^{2}N_{c}^{2}\rho _{B}^{2}}{4}%
+2g_{4}\left\langle \overline{q}q\right\rangle ^{2}\right) \Pi _{\pi
}(q^{2}=m_{\pi }^{*2})=0,  \label{mpi}
\end{equation}

\begin{equation}
f_{\pi }^{*}=N_{c}N_{f}g_{\pi qq}^{*}mI(q^{2}=m_{\pi }^{*2}),  \label{fpi}
\end{equation}
where the in-medium pion-quark-quark coupling constant, $g_{\pi qq}^{*}$, is
obtained by :

\begin{equation}
g_{\pi qq}^{*2}=\left[ \frac{d\Pi _{\pi }(q^{2})}{dq^{2}}\right]
_{q^{2}=m_{\pi }^{*2}}^{-1}.  \label{res}
\end{equation}

\section{Results}

We have six free parameters : the cutoff $\Lambda $, the bare quark mass $%
m_{0}$, and the coupling constants $g_{1}$, $g_{2}$, $g_{3}$ and $g_{4}$. As
usual, we have used for the fitting procedure the three following
constraints : the pion mass $m_{\pi }=135$ MeV, the pion decay constant $%
f_{\pi }=92.4$ MeV and the $\omega $ meson mass $m_{\omega }=782$ MeV in
vacuum. Since additional constraints like, for example, the value of the
quark condensate in vacuum, do not allow to determine the cutoff $\Lambda $
unambiguously, as often done\cite{bub}, we have considered several values
for $\Lambda $ or equivalently several values of the in-vacuum constituent
quark mass, $m_{vac}$, values ranging between $400$ MeV and $500$ MeV. Such
a rather large mass prevents the $\omega $ meson to be unstable against
decay into a quark-antiquark pair since $m_{\omega }^{*}$ is always lower
than $2m$ for every density. Note that smaller constituent quark masses can be obtained in NJL models which take the confinement into account by including Polyakov loops\cite{rtw} or using a confining interaction\cite{chs} in the Lagrangian. As already mentioned, in addition to these
in-vacuum constraints, we have also chosen to take into account recent
experimental results which provide constraints in the medium. In particular,
we have fixed $f_{\pi }^{*2}(\rho _{B}=\rho _{0})/f_{\pi }^{2}=0.64$ (where $%
\rho _{0}$ is the saturation density of nuclear matter) in accordance with
what is obtained in experiments on deeply bound pionic atoms\cite{suz}.
Moreover, the in-medium $\omega $ meson mass has been constrained to reproduce the experimental central value obtained either by the TAPS collaboration\cite{tap}, $%
m_{\omega }^{*}(\rho _{B}=0.6\rho _{0})=722$ MeV, or by the E325 experiment
at KEK\cite{nar}, $m_{\omega }^{*}(\rho _{B}=\rho _{0})=711$ MeV. Thus, two
families of parametrization sets denoted respectively by TAPS and KEK will be considered. Note
that once $m_{\pi }$, $f_{\pi }$, $m_{\omega }$, $f_{\pi }^{*}$, $m_{\omega
}^{*}$ and $m_{vac}$ are fixed, all the parameters are determined
unambiguously.

The results are shown on Fig.1 where we have plotted $m_{\omega
}^{*}/m_{\omega }$ as a function of $\left\langle \overline{q}q\right\rangle
/\left\langle \overline{q}q\right\rangle _{0}$ for the two parametrization
sets TAPS and KEK. Note that $\left\langle \overline{q%
}q\right\rangle /\left\langle \overline{q}q\right\rangle _{0}=0.8$
corresponds to a baryonic density close to the saturation one. The shaded
areas correspond to values of the $\omega $ meson mass for a constituent
quark mass ranging from 400 MeV to 500 MeV. As we can see, these areas
are rather narrow and the results are thus only weakly dependent on the
value of $m_{vac}$ used.

\begin{figure}[htb]
	\centering
		\epsfig{file=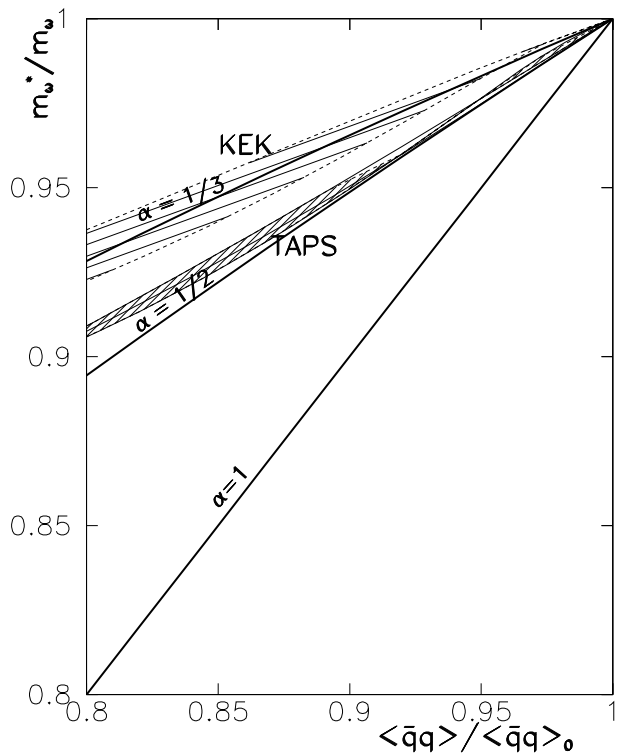,width=10cm}
	\caption{In-medium $\omega$ meson mass as a function of the quark condensate. The shaded areas correspond to values obtained for a constituent quark mass in the range $400<m_{vac}<500$ MeV. The full lines represent the scaling laws given by Eq.\ref{sca} for $%
\alpha =$1/3, 1/2 and 1.}
	\label{fig1}
\end{figure}

In order to determine an approximate form for the relation between the
vector meson mass and the quark condensate, we have considered scaling laws of
the general form : 
\begin{equation}
\frac{m_{\omega }^{*}}{m_{\omega }}=\left[ \left\langle \overline{q}%
q\right\rangle /\left\langle \overline{q}q\right\rangle _{0}\right] ^{\alpha
}.  \label{sca}
\end{equation}
Any value of $\alpha $ can be considered but we have chosen to show here the
results for $\alpha =1/2$ which corresponds to the Brown and Rho scaling and
for neighboring values: $\alpha =1/3$ and $\alpha =1$.

The full lines on Fig.1 represent the scaling laws
given by Eq.\ref{sca} for $\alpha =1/3$, $1/2$ and $1$. A rather good
agreement with the case $\alpha =1/2$ corresponding to the Brown and Rho
scaling law is obtained for the TAPS parametrization set while the KEK
result clearly favours $\alpha =1/3$. Assuming Eq.\ref{sca} for the scaling law, this result can be understood since to leading order $\left( \left\langle \overline{q}q\right\rangle
/\left\langle \overline{q}q\right\rangle _{0}\right) ^{\alpha }\simeq
1-\alpha \left( 1-\left\langle \overline{q}q\right\rangle /\left\langle 
\overline{q}q\right\rangle _{0}\right) $ and $\left\langle \overline{q}%
q\right\rangle /\left\langle \overline{q}q\right\rangle _{0}\simeq 1-\beta
\rho /\rho _{0}$ with $\beta =\sigma _{N}\rho _{0}/f_{\pi }^{2}m_{\pi }^{2}$
where $\sigma _{N}$ is the $\pi N$ sigma term. The TAPS or KEK results can
then be used to determine the product $\alpha \beta $ obtained by
eliminating the quark condensate in $m_{\omega }^{*}/m_{\omega }$. Using the
value $\sigma _{N}\simeq 35$ MeV obtained in the NJL model for $m\simeq 450$
MeV, the TAPS and KEK results provide respectively $\alpha$ close to $1/2$ and $1/3$ in agreement with our full calculation.

On the other hand, the experimental values of the in-medium $\omega $ meson mass are not
determined unambiguously. In particular, the TAPS collaboration obtained a rather large uncertainty including statistical and systematical errors. By taking into account such an uncertainty in our calculation, we have found that the scaling law (Eq.\ref{sca}) with $\alpha =1$ is clearly ruled out but the case $\alpha =1/3$ is not totally excluded. However, note that the central value of the TAPS result clearly favours $\alpha =1/2$.

\begin{figure}[htb]
	\centering
		\epsfig{file=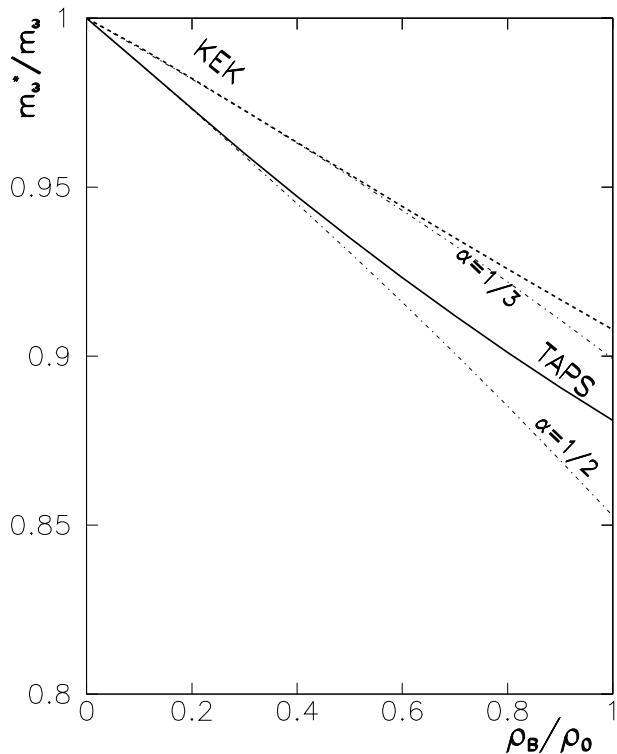,width=10cm}
	\caption{In-medium $\omega$ meson mass as a function of the baryonic density for the TAPS (solid curve) and KEK (dashed curve) parametrization sets. For comparison the scaling laws (\ref{sca}) with $\alpha = 1/2$ and $ 1/3 $ are also shown (dot-dashed curves).}
	\label{fig2}
\end{figure}

Let us now determine how the quark condensate and the $\omega $ meson mass
depend on the baryonic density for central values of the constituent
quark mass, i.e. for $m_{vac}=450$ MeV, respectively for the TAPS and KEK parametrizations.
The parameters of the NJL model are then $\Lambda =575$ MeV, $m_{0}=5.6$
MeV, $g_{1}\Lambda ^{2}=2.53$, $g_{2}\Lambda ^{2}=5.20$, $g_{3}\Lambda
^{8}=62.5$ and $g_{4}\Lambda ^{8}=2.27$ for the TAPS
parametrization and $\Lambda =575$ MeV, $m_{0}=5.6$ MeV, $g_{1}\Lambda
^{2}=2.58$, $g_{2}\Lambda ^{2}=4.12$, $g_{3}\Lambda ^{8}=12.4$ and $%
g_{4}\Lambda ^{8}=1.22$ for the KEK one. Let us mention that, whatever the
parametrization considered, the model provides a quark condensate in the
vacuum $\left\langle \overline{u}u\right\rangle ^{1/3}=-240$ MeV in good
agreement with the lattice calculations: $\left\langle \overline{u}%
u\right\rangle ^{1/3}=-(231\pm 4\pm 8\pm 6)$ MeV \cite{lat} and a critical
density close to four times the saturation one. On Fig.2, we have plotted $%
m_{\omega }^{*}/m_{\omega }$ (solid curves) for the two parametrizations
TAPS and KEK as well as $\left[ \left\langle \overline{q}q\right\rangle
/\left\langle \overline{q}q\right\rangle _{0}\right] ^{\alpha }$ for $\alpha
=1/2$ (dashed curve) and $\alpha =1/3$ (dot-dashed curve), as a function of
the dimensionless baryonic density $\rho _{B}/\rho _{0}$. As we can see, the
density dependences of the in-medium $\omega $ meson mass obtained with the
TAPS and KEK\ parametrizations lead to a drop close to 10\% at saturation
density (9\% for KEK and 12\% for TAPS). As already discussed, to a good level of approximation,
the in-medium $\omega $ meson mass determined using the KEK\ result varies as
a function of the baryonic density like the third root of the quark
condensate. On the other hand, using the TAPS result, the density dependence
of the in-medium $\omega $ meson mass is not very much different from the square root of the quark condensate reflecting the fact
that the results follow approximately the Brown and Rho scaling law up to
the saturation density.

\section{Conclusion}

We have determined the in-medium $\omega $ meson mass and quark condensate in a
NJL model with eight quark interaction terms. The parameters of this model have been determined using the meson properties in the vacuum but also in the medium through the value of the
pion decay constant obtained in experiments on deeply bound pionic atoms as
well as the $\omega $ meson mass measured either by the TAPS collaboration
or by the E325/KEK experiment. When the in-medium $\omega $ meson mass is
constrained to the experimental data obtained by the TAPS collaboration, the
Brown and Rho scaling law is approximately recovered. On the other hand,
when the KEK result is used, the in-medium $\omega $ meson mass varies rather
like the third root of the quark condensate. However, in both cases, this
corresponds to a drop of the $\omega $ meson mass at saturation density
close to 10\%, a result which is lower than those found in QCD sum rule
calculations where a decrease close to 15-25\% is generally obtained\cite
{hat,jin,kli}.

\ 

\end{document}